\documentclass[10pt]{article}
\usepackage{amsmath}
\usepackage{amsfonts}
\usepackage{amssymb}
\begin{document}

\author{E. G. Beltrametti$^\dagger$ and  S. Bugajski$^\ddagger$\footnote
{Supported by the Polish Committee for Scientific Research (KBN), grant No 7 
T11C 017 21}\\
$^\dagger$Department of Physics, University of Genoa and\\
Istituto Nazionale di Fisica Nucleare, Sezione di Genova\\
$^\ddagger$Institute of Physics, University of Silesia, and\\
Institute of Pure and Applied Informatics,\\
Polish Academy of Sciences
}
\date{}
\title{Entanglement and classical correlations in the quantum frame}
\maketitle
\begin{abstract}
The frame of classical probability theory can be generalized by enlarging the 
usual family of random variables in order to encompass nondeterministic ones: 
this leads to a frame in which two kinds of correlations emerge: the 
classical correlation which is coded in the mixed state of the physical 
system and a new 
correlation, to be called probabilistic entanglement, which may occur also at 
pure states. We examine to what extent this characterization of correlations 
can be applied to quantum mechanics. Explicit calculations on simple 
examples outline that a same quantum state can show only classical 
correlations or only entanglement depending on its statistical content; 
situations may also arise in which the two kinds of correlations 
compensate each other.
\end{abstract}
\section{Introduction}
The standard framework of classical statistical mechanics makes use of 
a convex set of states having the structure of a simplex, 
and adopts a family of observables, or random 
variables, which have a deterministic nature. More specifically, the 
states form the set $M_1^+(\Omega)$ of the probability 
measures on a measurable space $\Omega$ whose points -hence the Dirac 
measures on $\Omega$ to be denoted $\delta_{\omega},\:\omega\in \Omega$- 
represent the pure states. An observable taking 
values in a measurable space $\Xi$ corresponds to an affine map
$$
A: M_1^+(\Omega)\to M_1^+(\Xi),
$$
and the deterministic requirement is mirrored by the condition that $A$ has 
no dispersion on pure states, namely Dirac measures are mapped into 
Dirac measures, so that the observable $A$ becomes represented by a measurable 
function $\Omega\to\Xi$. Any two observables have a unique joint observable 
and a correlation between their outcomes can occur only at mixed states.

If the above framework is generalized by dropping out the deterministic 
requirement, so allowing also observables that have dispersion on pure states, 
we get a frame which has been discussed in [2,3,4,8]: the set of states is still 
the simplex $M_1^+(\Omega)$ but now the observables need not map Dirac 
measures on $\Omega$ into Dirac measures on the pertaining outcome space. 
Any two observables admit a joint observable but the latter is nonunique when 
the two observables have an indeterministic nature: a correlation between 
their outcomes is now relative to the considered joint observable and it need 
not vanish at pure states. In [5] it is shown that we can separate two
kinds of correlation: the {\it classical correlation}, which occurs only in 
a mixed state and is coded in the way the pure states are mixed up to get 
the nonpure state in question, and the {\it probabilistic entanglement} 
generated by the joint observable considered, and occurring also in pure states. 
Both correlations can be exhaustively characterized by corresponding density 
functions (correlation functions). As the name suggests, the probabilistic 
entanglement is analogous to the corresponding quantum concept.

In Section 2 we examine to what extent the notions of classical correlation 
and of probabilistic entanglement can be transferred to the quantum context: 
we will point at the fact the nonsimplex structure of the set of 
quantum states gives rise to ambiguities in separating classical correlation 
and entanglement when mixed states are considered. A quantum state can 
always be decomposed into a convex combination of pure states, so that it 
admits a representation in the set $M_1^+(\Omega_{\cal H})$ of the probability 
measures on the measurable space $\Omega_{\cal H}$ of the one-dimensional 
projectors of the 
Hilbert space $\cal H$, but this representation is in general nonunique. In 
fact, the observables adopted by quantum mechanics do 
not separate $M_1^+(\Omega_{\cal H})$: they define a partition 
of $M_1^+(\Omega_{\cal H})$ into equivalence classes that correspond to the 
density operators of $\cal H$. Recalling that mixed quantum states are 
associated with density operators of ${\cal H}$ and that pure states are 
associated with one-dimensional projectors, it is indeed well known that 
the convex decomposition of a density operator into 
one-dimensional projectors is nonunique. In Section 3 we will discuss 
a simple example which emphasizes the ambiguities said above.

The problem of defining the notion of entanglement and of classical 
correlation at mixed quantum states, 
and the related issue of characterizing the states that can (or cannot) 
exhibit entanglement, have received attention in the literature under 
different perspectives: as actual guides to the vast literature could serve 
[11] and [13]; we mention also [12] and the rigorous approach of 
Majewski [14] where some ideas similar to ours were formulated.

Let us recall that a quantum observable taking values in the measurable 
space $\Xi$ can be represented by a POV-measure 
$E:{\cal B}(\Xi) \to {\cal L}({\cal H})$ where ${\cal B}(\Xi)$ is 
$\sigma$-Boolean algebra of subsets of $\Xi$ and ${\cal L}({\cal H})$ is the 
family of the positive operators of $\cal H$ (see, {\it e.g.}, [9,10]). 
If $\Xi$ is the set of the reals 
then ${\cal B}(\Xi)$ is typically the Boolean algebra of the Borel subsets; if 
$\Xi$ is a finite set then ${\cal B}(\Xi)$ is typically the Boolean algebra of 
all subsets of $\Xi$.

Writing $S(\cal H)$ for the convex set of the density operators of $\cal H$, 
an equivalent representation of an observable taking values in $\Xi$ is 
given by an affine map $A$ of $S({\cal H})$ into 
the set $M_1^+(\Xi)$ of all probability measures on $\Xi$. For a given 
quantum state $D\in S({\cal H})$ the measure $A(D)$ is the outcome measure 
which is the result of a measurement of the observable $A$ on the state $D$. 
We write $E^{A}$ to denote the POV-measure associated to $A$ and recall that 
the probability of getting a value of the observable $A$ in the set 
$X\in{\cal B}(\Xi)$ at the state $D$ is given by the basic quantum rule
\begin{equation}
A(D)(X)=\mbox{Tr }(E^{A}(X)D).
\end{equation}
If the POV-measure $E^{A}$ reduces to a PV-measure on the real line, 
then the observable $A$ is known to correspond to a self-adjoint operator of 
${\cal H}$.

Notice that when $\Xi$ has the form of a Cartesian product $\Xi_1\times\Xi_2$ 
an observable $A:S({\cal H}) \to M_1^{+}(\Xi_1\times \Xi _2) $ 
defines two observables, $A_i:S({\cal H}) \to M_1^{+}(\Xi_i),\:\:i=1,2,$ by 
$A_i(D) :=\Pi_i( A(D)) $ where 
$\Pi_i:M_1^{+}(\Xi_1\times \Xi_2) \to M_1^{+}(\Xi_i)$ is the 
marginal projection.
The observable $A$ is then said to be a quantum joint observable of 
$A_1$ and $A_2.$ 
However, for a pair of observables $A_i:S({\cal H}) \to M_1^{+}(\Xi_i),$ 
$i=1,2,$ the existence of a quantum joint observable is not ensured.

\section{Correlations}
In probability theory a correlation between two parameter sets $\Xi_1$ 
and $\Xi _2$ is understood as a particular property of a probability 
measure $\nu $ on $\Xi_1\times \Xi_2,$ namely
$$
\nu \neq \nu_1\boxtimes \nu_2
$$
where $\nu_i=\Pi_i\nu $, $i=1,2$, is the marginal measure on $\Xi _i,$ and 
$\boxtimes $ stands for the product of measures. Thus, the notion of 
correlation just corresponds to the lack of independence (to a "mutual 
relationship", according to the Oxford Advanced Learner's Dictionary).

Consequently, we can say that a correlation between $\Xi_1$ and $\Xi_2$
encoded in $\nu \in M_1^{+}( \Xi_1\times \Xi_2) $ is what
distinguishes $\nu $ from $\nu_1\boxtimes \nu_2$. 
If we want to find a formal characterization of such a correlation 
we have to find how to describe the ''difference'' between 
$\nu $ and $\nu_1\boxtimes \nu_2.$ An exhaustive description of this 
"difference" is provided by the density function (the Radon-Nikodym 
derivative, see {\it e.g.} [1,7]) of $\nu $ w.r.t. $\nu_1\boxtimes \nu_2.$ Consequently, 
everything one can say about a correlation between $\Xi_1$ and $\Xi_2$ 
encoded in $\nu \in M_1^{+}( \Xi_1\times \Xi_2) $ is contained 
in the density function 
$$
\rho:=\frac{d\nu}{ d(\nu_1\boxtimes \nu_2)}
$$ 
which is a real-valued positive function on $\Xi_1\times \Xi_2$. The 
existence of this density function is ensured whenever $\nu$ is a discrete 
measure [5]. If in particular $\Xi_1$ and $\Xi_2$ are finite sets, the 
relationship between $\nu$ and $\nu_1\boxtimes \nu_2$ will take the form
$$
\nu(X)=\sum_{(\xi_1,\xi_2)\in X}\rho\cdot\nu_1\boxtimes\nu_2\:(\xi_1,\xi_2), 
\:\:\:\:\xi_1\in \Xi_1,\:\xi_2\in\Xi_2
$$
for every $X\subseteq\Xi_1\times\Xi_2$. In this case $\rho$ can be simply calculated 
by pointwise dividing the two measures $\nu$ and $\nu_1\boxtimes \nu_2$:
\begin{equation}
\rho(\xi_1,\xi_2)=\frac{\nu(\xi_1,\xi_2)}{\nu_1\boxtimes\nu_2\:(\xi_1,\xi_2)}\:.
\end{equation} 
The presence of a correlation is mirrored by the fact that $\rho$ is not the 
constant unit function.

We will be intersted in the case in which the two correlated sets 
$\Xi_1$ and $\Xi_2$ are value sets (sets of outcomes) of two observables, 
while the probability measure $\nu $ on $\Xi_1\times \Xi_2$ is the result 
of the measurement of a joint observable of them. 

In the standard context of the classical statistical mechanics, where the set 
of states is the simplex $M_1^+(\Omega)$ and only deterministic observables 
come into play, any two observables $A_1:M_1^+(\Omega)\to M_1^+(\Xi_1)$, 
$A_2:M_1^+(\Omega)\to M_1^+(\Xi_2)$ always admit the unique joint observable 
$A_1\boxtimes A_2$ defined by its action on the pure states
\begin{equation}
A_1\boxtimes A_2(\delta_{\omega}):=
A_1(\delta_{\omega})\boxtimes A_2(\delta_{\omega})\:\:\mbox{  for every }
\omega\in \Omega,
\end{equation}
and extended by affinity to the whole 
$M_1^+(\Omega)$. Thus, when we speak of a correlation between $A_1$ and $A_2$ 
at a state $\mu\in M_1^+(\Omega)$ the reference to the joint observable 
$A_1\boxtimes A_2$ is compulsory, and we have just to compare the two measures 
$A_1\boxtimes A_2(\mu)$ and $A_1(\mu)\boxtimes A_2(\mu)$. What we get is the 
classical correlation characterized by the density function (the Radon-Nicodym 
derivative)
\begin{equation}
\rho_c:=\frac{d(A_1\boxtimes A_2(\mu))}{ d(A_1(\mu)\boxtimes A_2(\mu))}\:.
\end{equation}
In view of Eq.(3) $\rho_c=1$ at pure states: a nontrivial classical correlation 
can appear only at mixed states.

If we go to the generalization of the standard classical frame by allowing also 
indeterministic observables, then the unicity of the joint observable breaks 
down: besides $A_1\boxtimes A_2$ other joint observables become possible (see 
[3,4,5]). When we speak of a correlation between $A_1$ and $A_2$ 
at a state $\mu\in M_1^+(\Omega)$ we have now to specify which joint 
observable $J(A_1,A_2)$ we refer to and we are naturally led to compare 
the two measures $J(A_1,A_2)(\mu)$ and $A_1(\mu)\boxtimes A_2(\mu)$, thus getting 
the correlation characterized by the density function
\begin{equation}
\rho_t:=\frac{d(J(A_1,A_2)(\mu))}{ d(A_1(\mu)\boxtimes A_2(\mu))}\:.
\end{equation}      
As discussed in [5] this correlation can be, in general, partitioned into two 
parts by first comparing the measure $J(A_1,A_2)(\mu)$ with 
$A_1\boxtimes A_2(\mu)$ and then comparing the measure 
$A_1\boxtimes A_2(\mu)$ with $A_1(\mu)\boxtimes A_2(\mu)$. The second step 
provides just the classical correlation said above, while the first step 
provides a correlation to be called {\it entanglement}. Clearly, the 
entanglement can emerge only when the 
joint observable referred to differs from $A_1\boxtimes A_2$, namely from 
the classical joint observable.
The density function associated to the entanglement will then be 
\begin{equation}
\rho_e:=\frac{d(J(A_1,A_2)(\mu))}{ d(A_1\boxtimes A_2(\mu))}\:,
\end{equation}       
and known properties of the Radon-Nicodym derivatives (see, {\it e.g.}, [1], 
Corollary 2.9.4, or [7], Sect. 32) give the product rule
\begin{equation}
\rho_t=\rho_c\cdot \rho_e\:.
\end{equation}
This motivates for $\rho_t$ the name of total correlation (hence the 
notation).  

Let us now come to the quantum frame, and consider two quantum 
observables $A_i:S( {\cal H}) \to M_1^{+}(\Xi_i)$, $i=1,2,$ admitting a joint 
observable: this is the case, for instance, when one deals with real valued 
observables represented by commuting self-adjoint operators 
(the joint observable is then unique). We can 
say that the two observables are correlated at the quantum state 
$D\in S({\cal H})$, relative to the 
given quantum joint observable $J(A_1,A_2):S( {\cal H})\to M_1^{+}(\Xi_1\times \Xi_2)$ 
iff 
$$ 
J(A_1,A_2)(D) \neq A_1(D)\boxtimes A_2(D). 
$$

The total correlation between the quantum observables $A_1$ and $A_2$ 
relative to the joint observable $A$ (at the state $D$) is then exhaustively 
described, as in Eq.(5), by the total correlation function 
\begin{equation}
\rho_t=\frac{d(J(A_1,A_2)(D)) }{d( A_1(D) \boxtimes A_2(D))}\:. 
\end{equation}
But if we tackle the problem of separating the classical correlation and 
the entanglement then we are faced with the translation to the quantum frame 
of the product $A_1\boxtimes A_2$. We can mirror Eq.(3) by defining 
$A_1\boxtimes A_2$ on the pure states according to
\begin{equation}
A_1\boxtimes A_2 (P):=A_1 (P)\boxtimes A_2 (P)
\end{equation}
for every one-dimensional projector $P$ of ${\cal H}$. But the extension by 
affinity to the whole set of quantum states $S({\cal H})$ makes sense only if 
we refer to a specific convex decomposition into pure states of the mixed state 
$D$, and this decomposition is known to be nonunique. In other words, if the 
mixed state (density operator) $D$ under discussion admits the convex 
decomposition into pure states
\begin{equation}
D=\sum_i w_iP_i
\end{equation}
where the $w_i$'s are positive numbers whose sum is 1 and the $P_i$'s are 
one-dimensional projectors, then we can affinely define $A_1\boxtimes A_2$ 
on the r.h.s. of Eq.(10) getting the measure
\begin{equation}
\sum_i w_i A_1\boxtimes A_2 (P_i)=\sum_i w_i A_1 (P_i)\boxtimes A_2 (P_i),
\end{equation}
but this measure is not invariant under different choices of the convex 
decomposition of $D$.

In view of the above fact one can speak of classical correlation and 
of entanglement in the quantum context only with reference to a given 
convex decomposition of the (mixed) state under discussion. The corresponding 
density functions will read (see Eqs.(4),(6))
\begin{equation}
\rho_c:=\frac{d(\sum_i w_i A_1 (P_i)\boxtimes A_2 (P_i))}{d(A_1D\boxtimes A_2D)}
\end{equation}
and
\begin{equation}
\rho_e:=\frac{d(J(A_1,A_2)(D))}{d(\sum_i w_i A_1 (P_i)\boxtimes A_2 (P_i))}\:.
\end{equation}
Let us stress that only the product $\rho_c \cdot \rho_e$ which equals 
$\rho_t$ (see Eq.(7)) has the property of being invariant under different 
convex decompositions of the quantum state $D$, while neither $\rho_c$ nor 
$\rho_e$ have such an invariance.

\section{A two-qubit example}

We will illustrate the introduced concepts on a simple quantum-mechanical 
example based on a Hilbert space of the form ${\cal H}={\bf C}^2\otimes 
{\bf C}^2$: it can be viewed as the composition of two spin-$\frac{1}{2}$ 
or as a two-qubit system.

Let $\{\psi_+,\psi_-\}$ be an orthonormal basis of ${\bf C}^2$ and let 
$P_{+},P_{-}$ be the corresponding (one-dimensional) projectors. A canonical 
orthonormal basis of ${\bf C}^2\otimes {\bf C}^2$ is provided by 
$\{\psi_{+}\otimes \psi_{+}\:,\:\psi_{-}\otimes \psi_{-}\:,\:
\psi_{+}\otimes \psi_{-}\:,\:\psi_{-}\otimes \psi_{+}\}$ and the associated 
one-dimensional projectors read:
$$
P_{++}=P_{+}\otimes P_{+}\:,\:\:P_{--}=P_{-}\otimes P_{-}\:,\:\:
P_{+-}=P_{+}\otimes P_{-}\:,\:\:P_{-+}=P_{-}\otimes P_{+}\:.
$$

Consider the two observables $A_i:S({\bf C}^2\otimes {\bf C}^2)\to 
M_1^+(\{\frac{1}{2},-\frac{1}{2}\})$, $i=1,2$, described by the self-adjoint 
operators on ${\bf C}^2\otimes {\bf C}^2$:
$$
\widehat{A}_1 \:=\:(\frac {1}{2}P_{+}-\frac {1}{2}P_{-})\otimes I,
\:\:\:\:
\widehat{A}_2 \:=\:I\otimes (\frac {1}{2}P_{+}-\frac {1}{2}P_{-}),
$$
where $I$ denotes the identity operator in ${\bf C}^2$. 
We can view $A_i$ as the observable describing the z-component 
of the spin of the $i$-th subsystem. According to quantum mechanics the only 
admissible joint observable of $A_1,A_2 $ is the observable 
$J(A_1,A_2):S({\bf C}^2\otimes{\bf C}^2)\to 
M_1^+(\{\frac{1}{2},-\frac{1}{2}\}\times \{\frac{1}{2},-\frac{1}{2}\})$ 
which corresponds to the PV measure $E^{J(A_1,A_2)}$ defined by
$$ 
E^{J(A_1,A_2)}\mbox{$(\frac{1}{2},\frac{1}{2})$} =P_{++}\:,\:\:\:\:\:
E^{J(A_1,A_2)}\mbox{$( -\frac{1}{2},-\frac{1}{2})$} =P_{--}\:,
$$
$$
E^{J(A_1,A_2)}\mbox{$(\frac{1}{2},-\frac{1}{2})$} =P_{+-}\:,\:\:\:\:\:
E^{J(A_1,A_2)}\mbox{$(-\frac{1}{2},\frac{1}{2})$} =P_{-+}\:.
$$

We will be concerned with the correlation between $A_1$ and $A_2$ relative to 
the joint observable $J(A_1,A_2)$ at various quantum states: this will point at the 
fact that, in the quantum frame, the splitting of the correlation into 
a classical part and an entanglement might become a matter of convention.

In the sequel we will have to compute the various measures involved in the 
correlations of interest: to do that we will refer to Eq.(1), noticing that 
the trace is linear and that for a pure state, say $P$, the r.h.s. of Eq.(1) 
takes the form $(\phi,E^A(X)\phi)$ where $\phi$ is any unit vector in the 
one-dimensional subspace onto which $P$ projects. 
We will write $\eta_{\frac{1}{2}}$ to denote the Dirac measure 
on $\{\frac{1}{2},-\frac{1}{2}\}$ concentrated at the value $\frac{1}{2}$ 
(similarly for $\eta_{-\frac{1}{2}}$), and $\eta_{(\frac{1}{2},\frac{1}{2})}$ 
to denote the Dirac measure on $\{\frac{1}{2},-\frac{1}{2}\}\times 
\{\frac{1}{2},-\frac{1}{2}\}$ concentrated at the point 
$(\frac{1}{2},\frac{1}{2})$ (similarly for $\eta_{(\frac{1}{2},-\frac{1}{2})}$, 
$\eta_{(-\frac{1}{2},\frac{1}{2})}$, $\eta_{(-\frac{1}{2},-\frac{1}{2})}$).

\medskip

(i) {\bf Separable mixed state}

A {\it separable mixed state} is represented by a density operator which 
decomposes into the convex combination of pure product states. A canonical 
example is provided by the density operator
\begin{equation}
D=w_1P_{++}+w_2P_{--}+w_3P_{+-}+w_4P_{-+}\:.
\end{equation}

In order to get the total correlation we have now to compare the two 
measures $J(A_1,A_2)(D)$ and $A_1(D)\boxtimes A_2(D)$.
 
The measure $J(A_1,A_2)(D)$ on $\{\frac{1}{2},-\frac{1}{2}\}\times 
\{\frac{1}{2},-\frac{1}{2}\}$ is easily obtained looking 
at the explicit expression of $E^{J(A_1,A_2)}$ given above. We get
$$
J(A_1,A_2)(D)=w_1 \eta_{(\frac{1}{2},\frac{1}{2})}+
w_2\eta_{(-\frac{1}{2},-\frac{1}{2})}+w_3 \eta_{(\frac{1}{2},-\frac{1}{2})}+ 
w_4\eta_{(-\frac{1}{2},\frac{1}{2})}\:.
$$

The measure $A_1(D)$ on $\{\frac{1}{2},-\frac{1}{2}\}$ can be obtained by 
an analogous procedure: noticing that $E^{A_1}(\frac{1}{2})=P_{+}\otimes 
I$ and $E^{A_1}(-\frac{1}{2})=P_{-}\otimes I$ we get, 
$$
A_1(D)=(w_1+w_3)\eta_{\frac{1}{2}}+(w_2+w_4)\eta_{-\frac{1}{2}}\:.
$$
Similarly we have
$$
A_2(D)=(w_1+w_4)\eta_{\frac{1}{2}}+(w_2+w_3)\eta_{-\frac{1}{2}}\:.
$$

Hence the product measure $A_1(D)\boxtimes A_2(D)$ takes the form 
\begin{eqnarray}
A_1(D)\boxtimes A_2(D)&=&
(w_1+w_3)(w_1+w_4) \eta_{(\frac{1}{2},\frac{1}{2})}+
(w_2+w_4)(w_2+w_3) \eta_{(-\frac{1}{2},-\frac{1}{2})} \nonumber
\\
&+&(w_1+w_3)(w_2+w_3)\eta_{(\frac{1}{2},-\frac{1}{2})}+ 
(w_2+w_4)(w_1+w_4)\eta_{(-\frac{1}{2},\frac{1}{2})}. \nonumber
\end{eqnarray}

The density function of the total correlation between $A_1$ and $A_2$ at 
the state $D$ will then turn out to be (see Eqs.(2),(8)):
$$
\rho_t\mbox{$( \frac{1}{2},\frac{1}{2})$}=\frac{w_1}
{(w_1+w_3)(w_1+w_4)}\:,
\:\:\:\:\:
\rho_t\mbox{$(-\frac{1}{2},-\frac{1}{2})$}=\frac{w_2}
{(w_2+w_4)(w_2+w_3)}\:,
$$
$$
\rho_t\mbox{$( \frac{1}{2},-\frac{1}{2})$}=\frac{w_3}
{(w_1+w_3)(w_2+w_3)}\:,
\:\:\:\:\:
\rho_t\mbox{$(-\frac{1}{2},\frac{1}{2})$}=\frac{w_4}
{(w_2+w_4)(w_1+w_4)}\:.
$$

In order to examine how this total correlation splits into classical 
correlation and entanglement we have now to evaluate the measure 
(see Eqs.(9),(11)) 
$$
w_1A_1\boxtimes A_2 (P_{++})+w_2 A_1\boxtimes A_2 (P_{--})
+w_3A_1\boxtimes A_2(P_{+-})+w_4A_1\boxtimes A_2(P_{-+}) 
$$
where $A_1\boxtimes A_2 (P_{++})=A_1(P_{++})\boxtimes A_2 (P_{++})$ and so on. 
The calculation goes as before: for instance we have $A_1(P_{++})=A_2(P_{++})=
\eta_{\frac{1}{2}}$ so that 
$A_1(P_{++})\boxtimes A_2 (P_{++})=\eta_{(\frac{1}{2},\frac{1}{2})}$, and 
similarly for the other terms. The result 
is that the measure above equals exactly the measure $J(A_1,A_2)(D)$ said before. 
This means that the density function $\rho_c$ of the classical correlation 
coincides with $\rho_t$ (see Eq.(12)) while the density function $\rho_e$ 
(see Eq.(13)) is the constant unit function. 
In other words, the total correlation between $A_1$ and $A_2$ 
at the state $D$ appears to be 
entirely a classical correlation, without any entanglement coming into play.  

In the Appendix we will prove that the absence of entanglement holds 
true also for every bipartite separable mixed state.

\medskip

(ii) {\bf Bell diagonal state}

Instead of the canonical basis 
$\{\psi_{+}\otimes \psi_{+},\:\psi_{-}\otimes \psi_{-},\:
\psi_{+}\otimes \psi_{-},\:\psi_{-}\otimes \psi_{+}\}$ used before, let us now 
turn to the Bell basis
$$
\Phi_1:=\frac{1}{\sqrt{2}}(\psi_{+}\otimes \psi_{+}+\psi_{-}\otimes \psi_{-}),
\:\:\:\:\:
\Phi_2:=\frac{1}{\sqrt{2}}(\psi_{+}\otimes \psi_{+}-\psi_{-}\otimes \psi_{-}),
$$
$$
\Phi_3:=\frac{1}{\sqrt{2}}(\psi_{+}\otimes \psi_{-}+\psi_{-}\otimes \psi_{+}),
\:\:\:\:\:
\Phi_4:=\frac{1}{\sqrt{2}}(\psi_{+}\otimes \psi_{-}-\psi_{-}\otimes \psi_{+}),
$$
and let $P_1,\:P_2,\:P_3,\:P_4$ be the 
corresponding one-dimensional projectors.

A convex combination of the form 
\begin{equation}
D^{\prime}=w_1^{\prime}P_1+w_2^{\prime}P_2+
w_3^{\prime}P_3+w_4^{\prime}P_4
\end{equation}
is called a {\it Bell diagonal state} [6]. We are going to obtain the 
correlation functions for the observables $A_1,\:A_2$ at such a state.

The outcome measure $J(A_1,A_2)(D^{\prime})$ of the quantum joint observable of 
$A_1$ and $A_2$ is easily found to be
$$
J(A_1,A_2)(D^{\prime})=
\frac{1}{2}(w_1^{\prime}+w_2^{\prime})
(\eta_{(\frac{1}{2},\frac{1}{2})}+\eta_{(-\frac{1}{2},-\frac{1}{2})})+
\frac{1}{2}(w_3^{\prime}+w_4^{\prime})
(\eta_{(\frac{1}{2},-\frac{1}{2})}+\eta_{(-\frac{1}{2},\frac{1}{2})}).
$$

The two measures $A_1(D^{\prime})$ and $A_2(D^{\prime})$ have the uniform 
structure $\frac{1}{2}\eta_{\frac{1}{2}}+\frac{1}{2}\eta_{-\frac{1}{2}}$ 
so that also their product is uniformly distributed over the four-point space
$\{\frac{1}{2},-\frac{1}{2}\}\times\{\frac{1}{2},-\frac{1}{2}\}$:
$$
A_1(D^{\prime}) \boxtimes A_2(D^{\prime})=
\frac{1}{4}(\eta_{(\frac{1}{2},\frac{1}{2})}+
\eta_{(-\frac{1}{2},-\frac{1}{2})}+
\eta_{(\frac{1}{2},-\frac{1}{2})}+\eta_{(-\frac{1}{2},\frac{1}{2})}).
$$

Therefore, the density function of the total correlation between $A_1$ and 
$A_2$ at the Bell state $D^{\prime}$ is
$$
\rho_t^{\prime}\mbox{$( \frac{1}{2},\frac{1}{2})$}=
\rho_t^{\prime}\mbox{$(-\frac{1}{2},-\frac{1}{2})$}=
2(w_1^{\prime}+w_2^{\prime})\:\:,\:\:\:\:\: 
\rho_t^{\prime}\mbox{$(\frac{1}{2},-\frac{1}{2})$}=
\rho_t^{\prime}\mbox{$(-\frac{1}{2},\frac{1}{2})$}=
2(w_3^{\prime}+w_4^{\prime}).
$$ 

In order to see how this total correlation could be separated into classical 
correlation and entanglement we must go to the measure (see Eq.(11)) 
$$
w_1^{\prime}A_1\boxtimes A_2(P_1)+w_2^{\prime}A_1\boxtimes A_2(P_2)+
w_3^{\prime}A_1\boxtimes A_2(P_3)+w_4^{\prime}A_1\boxtimes A_2(P_4),
$$
which is easily seen to coincide with the uniform product measure 
$A_1(D^{\prime}) \boxtimes A_2(D^{\prime})$. Therefore, by inspection of 
Eqs.(12),(13), we conclude that $\rho_e^{\prime}=\rho_t^{\prime}$ while 
$\rho_c^{\prime}$ is the constant unit function. In other words, the 
correlation between $A_1$ and $A_2$ at the Bell state $D^{\prime}$ appears 
to be entirely an entanglement without any classical correlation coming 
into play.  

The absence of any classical correlation that we have found seems to disagree 
with the result of [12], where a numerical measure for classical correlation 
is introduced which does not vanish at some Bell diagonal state. This might 
point at the fact that such a numerical measure does not fully capture our 
notion of classical correlation.

A somewhat similar disagreement with previous literature occurs also when we 
look at the entanglement density function $\rho_e^{\prime}$ 
(=$\rho_t^{\prime}$) given above which is nonconstant whenever 
$w_1^{\prime}+w_2^{\prime}\neq w_3^{\prime}+w_4^{\prime}$. Indeed, according 
to [6] and [13] a Bell diagonal state shows entanglement only if one of the weights 
$w_1^{\prime}, \: w_2^{\prime},\: w_3^{\prime},\: w_4^{\prime}$ (in our 
notations) is bigger than 
$\frac{1}{2}$: clearly, this would imply the inequality 
$(w_1^{\prime}+w_2^{\prime}\neq w_3^{\prime}+w_4^{\prime}$ but the reverse 
implication does not hold. Again, this might point at the fact that the 
numerical measure of entanglement introduced in [6] (the "entanglement 
of formation") does not cover exactly our definition of entanglement.
 
\medskip

(iii) {\bf A degenerate state}

As far as the density operators $D$ and $D^{\prime}$ considered in items (i) 
and (ii) have no degenerate eigenvalues, 
that is as far as the $w_i$'s and the $w_i^{\prime}$'s in Eqs.(14) and (15) 
are pairwise distinct, 
it is guaranteed that $D$ and $D^{\prime}$ represent distinct quantum states. 
In this case the fact that the correlation between $A_1$ and $A_2$ at the state 
$D$ is purely classical while at the state $D^{\prime}$ it is just entanglement 
makes no problem. But a peculiar feature emerges when we consider 
degenerate eigenvalues, for instance when we assume
$$
w_1=w_2=w_1^{\prime}=w_2^{\prime}=a\:\:\: \mbox{ and }\:\:\: 
w_3=w_4=w_3^{\prime}=w_4^{\prime}=b 
$$ 
with $a+b=\frac{1}{2}$. In this case $D$ and $D^{\prime}$ actually 
represent the same quantum state, say $D_d$, 
since the only difference among their convex decompositions is a different 
choice of an orthonormal basis within the degenerate eigenspaces.

As expected we have now $\rho_t=\rho_t^{\prime}$, explicitly:
$$
\rho_t\mbox{$(\frac{1}{2},\frac{1}{2})$}=
\rho_t^{\prime}\mbox{$( \frac{1}{2},\frac{1}{2})$}=
\rho_t(\mbox{$-\frac{1}{2},-\frac{1}{2})$}=
\rho_t^{\prime}\mbox{$(-\frac{1}{2},-\frac{1}{2})$}=4a,
$$
$$
\rho_t\mbox{$(\frac{1}{2},-\frac{1}{2})$}=
\rho_t^{\prime}\mbox{$(\frac{1}{2},-\frac{1}{2})$}=
\rho_t\mbox{$(-\frac{1}{2},\frac{1}{2})$}=
\rho_t^{\prime}\mbox{$(-\frac{1}{2},\frac{1}{2})$}=4b.
$$ 

But, according to the results of items (i) and (ii), we have now that this 
correlation appears to be entirely a classical correlation if we refer to 
the convex combination
\begin{equation}
aP_{++}+aP_{--}+bP_{+-}+bP_{-+}
\end{equation}
while it appears to be entirely an entanglement if we refer to the convex 
combination
\begin{equation}
aP_1+aP_2+bP_3+bP_4\:,
\end{equation}
despite the fact that these two convex combinations correspond to the same 
quantum state.

This result emphasizes the fact that the separation of entanglement and 
classical correlation is possible only if we know the {\it statistical content} 
of the quantum mixed state, {\it i.e.} the actual decomposition of the mixed 
state into a convex combination of pure states, but this {\it statistical 
content} is in general not uniquely specified by the von Neumann description of 
quantum mixed states.

Let us further remark that also the convex combination
\begin{equation}
aP_{++}+aP_{--}+bP_3+bP_4
\end{equation}
represents the same quantum state $D_d$ expressed by Eq.(16) or by Eq.(17). 
Clearly, the total corelation between $A_1$ and $A_2$ at 
this new convex combination is the same as before, but now the density 
functions of the classical correlation and of the entanglement turn out to be 
$$
\rho_c\mbox{$( \frac{1}{2},\frac{1}{2})$}=
\rho_c\mbox{$(-\frac{1}{2},-\frac{1}{2})$}=2(2a+b)\:,\:\:\:\:\:
\rho_c\mbox{$( \frac{1}{2},-\frac{1}{2})$}=
\rho_c\mbox{$(-\frac{1}{2},\frac{1}{2})$}=2b\:,
$$
and
$$
\rho_e\mbox{$( \frac{1}{2},\frac{1}{2})$}=
\rho_e\mbox{$(-\frac{1}{2},-\frac{1}{2})$}=\frac{2a}{2a+b}\:,\:\:\:\:\:
\rho_e\mbox{$( \frac{1}{2},-\frac{1}{2})$}=
\rho_e\mbox{$( -\frac{1}{2},\frac{1}{2})$}=2\: .
$$
Thus, if one refers to the convex decomposition of 
Eq.(18), the total correlation appears to be partially a classical correlation 
and partially an entanglement. 

In the totally degenerate case 
$a=b\:(=\frac{1}{4})$, hence in the case of the "most mixed" state, the 
density function of the total correlation is, as expected, 
the constant unit function, no matter which convex decomposition one refers to. 
But, if the convex decomposition 
of Eq.(18) is referred to, there is still some classical correlation 
and some entanglement: indeed we find
$$
\rho_c\mbox{$( \frac{1}{2},\frac{1}{2})$}=
\rho_c\mbox{$( -\frac{1}{2},-\frac{1}{2})$}=\frac{3}{2}\:,\:\:\:\:\:\:
\rho_c\mbox{$( \frac{1}{2},-\frac{1}{2})$}=
\rho_c\mbox{$( -\frac{1}{2},\frac{1}{2})$}=\frac{1}{2}\:,
$$
while
$$
\rho_e\mbox{$(\frac{1}{2},\frac{1}{2})$}=
\rho_e\mbox{$(-\frac{1}{2},-\frac{1}{2})$}=\frac{2}{3}\:,\:\:\:\:\:\:
\rho_e\mbox{$(\frac{1}{2},-\frac{1}{2})$}=
\rho_e\mbox{$(-\frac{1}{2},\frac{1}{2})$}=2\:,
$$
in agreement with the product rule $\rho_c\cdot\rho_e=\rho_t$.
A classical correlation and an entanglement survive at the mixture of Eq.(18)
even if $a=b$, though in absence of a total correlation. This example 
shows a new and unexpected effect: even if a state shows no total correlation 
at all, one can find both classical and quantum correlations that 
compensate each other.

\section*{Appendix}
We refer to the two-qubit example of Section 3, and consider the observables 
$A_1,A_2$ there defined representing the z-component of the spin of the two 
subsystems. Again $J(A_1,A_2)$ denotes their joint observable and 
$E^{J(A_1,A_2)}$ is the corresponding PV measure. We are going to show that 
there is no entanglement between $A_1$ and $A_2$ at any bipartite separable 
mixed state represented by the density operator 
$$
D=\sum_i w_i{\cal P}_i\otimes {\cal Q}_i
$$
where ${\cal P}_i$ and ${\cal Q}_i$ are one-dimensional projectors of 
${\bf C}^2$, with $i$ ranging over the positive integers (actually what we 
are going to show easily generalizes to integrals of product pure states with 
respect to arbitrary probability measures over pure states of the two-qubit 
system).

In order to calculate the entanglement function we have to compare the two 
measures $J(A_1,A_2)(D)$ and $\sum_i w_i A_1({\cal P}_i\otimes {\cal Q}_i)
\boxtimes A_2({\cal P}_i\otimes {\cal Q}_i)$.

Looking at the explicit expression of $E^{J(A_1.A_2)}$ given in Section 3, 
and recalling that $\psi_+, \psi_-$ denote the orthonormal vectors of 
${\bf C}^2$ representing the spin-up and spin-down states along the z-axis, we 
obtain for $J(A_1,A_2)(D)$ the explicit form
\begin{eqnarray}
\sum_iw_i(\psi_+,{\cal P}_i\psi_+)(\psi_+,{\cal Q}_i\psi_+)
\eta_{(\frac{1}{2},\frac{1}{2})}+
\sum_iw_i(\psi_-,{\cal P}_i\psi_-)(\psi_-,{\cal Q}_i\psi_-)
\eta_{(-\frac{1}{2},-\frac{1}{2})}&+& \nonumber
\\
\sum_iw_i(\psi_+,{\cal P}_i\psi_+)(\psi_-,{\cal Q}_i\psi_-)
\eta_{(\frac{1}{2},-\frac{1}{2})}+
\sum_iw_i(\psi_-,{\cal P}_i\psi_-)(\psi_+,{\cal Q}_i\psi_+)
\eta_{(-\frac{1}{2},\frac{1}{2})}&.& \nonumber
\end{eqnarray}
In fact, looking for instance at the first term of the above expression, 
we have just to refer to Eq.(1) and recall that at the 
point $\{\frac{1}{2},\frac{1}{2}\}$ 
the PV measure $E^{J(A_1,A_2)}$ takes the value $P_{++}$, {\it i.e.}, the 
projector onto the state $\psi_+\otimes \psi_+$, so that the value of 
$J(A_1,A_2)(D)$ at that point becomes
$$
\sum_i w_i\mbox{Tr}(P_{++}\cdot{\cal P}_i\otimes{\cal Q}_i)=
\sum_i w_i (\psi_+\otimes \psi_+, {\cal P}_i\otimes{\cal Q}_i\:  
\psi_+\otimes \psi_+
$$
(with similar remarks applying for the other terms).

On the other hand we have
$$
A_1( {\cal P}_i\otimes{\cal Q}_i)=(\psi_+,{\cal P}_i\psi_+)\eta_{\frac{1}{2}}+
(\psi_-,{\cal P}_i\psi_-)\eta_{-\frac{1}{2}}
$$
and
$$
A_2( {\cal P}_i\otimes{\cal Q}_i)=(\psi_+,{\cal Q}_i\psi_+)\eta_{\frac{1}{2}}+
(\psi_-,{\cal Q}_i\psi_-)\eta_{-\frac{1}{2}}
$$
as one sees by noticing, for instance, that at the point $\{\frac{1}{2}\}$ 
the PV measure $E^{A_1}$ takes 
the value $P_+\otimes I$ (see Section 3), so that 
at this point the value of $A_1( {\cal P}_i\otimes{\cal Q}_i)$ becomes 
(see Eq.(1)) Tr($P_+\otimes I\cdot {\cal P}_i\otimes{\cal Q}_i$)=
Tr($P_+{\cal P}_i$)=($\psi_+,{\cal P}_i\psi_+$); and similarly for the other 
terms. Hence the mixture $\sum_i w_i A_1( {\cal P}_i\otimes{\cal Q}_i)
\boxtimes A_2( {\cal P}_i\otimes{\cal Q}_i)$ of the product measures is 
immediately seen to reproduce exactly the measure $J(A_1,A_2)(D)$, so that 
we conclude that the entanglement density function $\rho_e$ is indeed the 
constant unit function.

As expected, there can be classical correlations between the two observables 
$A_1,A_2$ at the state $D$. In fact we have
$$
A_1(D)=\sum_i w_i (\psi_+,{\cal P}_i\psi_+)\eta_{\frac{1}{2}}+
\sum_i w_i(\psi_-,{\cal P}_i\psi_-)\eta_{-\frac{1}{2}}
$$
and
$$
A_2(D)=\sum_i w_i (\psi_+,{\cal Q}_i\psi_+)\eta_{\frac{1}{2}}+
\sum_i w_i(\psi_-,{\cal Q}_i\psi_-)\eta_{-\frac{1}{2}},
$$
so that the product measure takes the form
\begin{eqnarray}
A_1(D)\boxtimes A_2(D)&=&
(\sum_i w_i (\psi_+,{\cal P}_i\psi_+)\:(\sum_i w_i (\psi_+,{\cal Q}_i\psi_+)
\:\eta_{(\frac{1}{2},\frac{1}{2})} \nonumber
\\
&+&(\sum_i w_i (\psi_-,{\cal P}_i\psi_-)\:(\sum_i w_i (\psi_-,{\cal Q}_i\psi_-)
\:\eta_{(-\frac{1}{2},-\frac{1}{2})} \nonumber
\\
&+&(\sum_i w_i (\psi_+,{\cal P}_i\psi_+)\:(\sum_i w_i (\psi_-,{\cal Q}_i\psi_-)
\:\eta_{(\frac{1}{2},-\frac{1}{2})} \nonumber
\\
&+&(\sum_i w_i (\psi_-,{\cal P}_i\psi_-)\:(\sum_i w_i (\psi_+,{\cal Q}_i\psi_+)
\:\eta_{(-\frac{1}{2},\frac{1}{2})}. \nonumber
\end{eqnarray}
Clearly this measure does not coincide, in general, with $J(A_1,A_2)(D)$ so 
that we get a nontrivial density function $\rho_c$: the considered spin 
observables $A_1,A_2$ can exhibit classical correlations at $D$. 

To exemplify the above results, let us consider the particular case
$$
D=w P_+\otimes P_+ +(1-w)P_x\otimes P_x
$$
where $P_+$ is the the projector on the spin-up state $\psi_+$ along the 
z-axis while $P_x$ is the projector on the eigenstate of the x-component of 
the spin corresponding to the eigenvalue $+\frac{1}{2}$.

Noticing that $(\psi_+,P_x\psi_+)=(\psi_-,P_x\psi_-)=\frac{1}{2}$ we see, by 
inspection of the previous formulas, that both the measures $J(A_1,A_2)(D)$ 
and $w A_1(P_+\otimes P_+)\boxtimes  A_2(P_+\otimes P_+)+
(1-w) A_1(P_x\otimes P_x)\boxtimes  A_2(P_x\otimes P_x)$ now take the form
$$
\frac{1+3w}{4}\eta_{(\frac{1}{2},\frac{1}{2})}+
\frac{1-w}{4}\eta_{(-\frac{1}{2},-\frac{1}{2})}+
\frac{1-w}{4}\eta_{(\frac{1}{2},-\frac{1}{2})}+
\frac{1-w}{4}\eta_{(-\frac{1}{2},\frac{1}{2})},
$$
leaving no room for entanglement.

On the other hand the product measure $A_1(D)\boxtimes A_2(D)$ now reads
$$
\frac{(1+w)^2}{4}\eta_{(\frac{1}{2},\frac{1}{2})}+
\frac{(1-w)^2}{4}\eta_{(-\frac{1}{2},-\frac{1}{2})}+
\frac{1-w^2}{4}\eta_{(\frac{1}{2},-\frac{1}{2})}+
\frac{1-w^2}{4}\eta_{(-\frac{1}{2},\frac{1}{2})},
$$
so that the classical correlation function becomes, for $w\neq 1$,
\begin{eqnarray}
\rho_c\mbox{$(\frac{1}{2},\frac{1}{2})$}=\frac{1+3w}{(1+w)^2}&,&\:\:\:
\rho_c\mbox{$(-\frac{1}{2},-\frac{1}{2})$})=\frac{1}{1-w}\:\:,
\nonumber
\\
\rho_c\mbox{$(\frac{1}{2},-\frac{1}{2})$}=\frac{1}{1+w}&,&\:\:\:
\rho_c\mbox{$(-\frac{1}{2},\frac{1}{2})$}=\frac{1}{1+w}\:\:,\nonumber
\end{eqnarray}
while for $w=1$ both $J(A_1,A_2)(D)$ and $A_1(D)\boxtimes A_2(D)$ become 
concentrated at the point $\{\frac{1}{2},\frac{1}{2}\}$.

\section*{References}

\noindent[1] H, Bauer, {\it Probability Theory and Elements of Measure Theory}, 
Academic Press, London, 1981

\smallskip

\noindent[2]  E. G. Beltrametti and S. Bugajski, J. Phys. A: Math.Gen. {\bf 28} 
(1995) 3329

\smallskip

\noindent[3] S. Bugajski, Int. J. Theor. Phys. {\bf 35} (1996) 2229 

\smallskip

\noindent[4]  E. G. Beltrametti and S. Bugajski, J. Phys. A: Math.Gen. {\bf 29} 
(1996) 247

\smallskip

\noindent[5] E. G. Beltrametti and S. Bugajski, {\it Correlations and 
entanglement in probability theory}, arXiv:quant-ph/0211083, 14 Nov. 2002

\smallskip

\noindent[6]  C. H. Bennet, D. P. Di Vincenzo, J. A. Smolin, W. K. Wootters, 
Phys. Rev. A {\bf 54} (1996) 3824

\smallskip

\noindent[7] P. Billingsley, {\it Probability and Measure}, Wiley, 
New York, 1979

\smallskip

\noindent[8] S. Bugajski, Mathematica Slovaca {\bf 51} (2001) 321 and 
343

\smallskip

\noindent[9] P. Busch, M. Grabowski and P.J. Lahti, {\it Operational Quantum 
Physics}. Lecture Notes in Physics {\bf m 31}, 
2nd Edition, Springer-Verlag, Berlin, 1995

\smallskip

\noindent[10] P. Busch, P. J. Lahti and P. Mittelstaedt, {\it The 
Quantum Theory of Measurement}, Lecture Notes in Physics {\bf m 2}, 
2nd Edition, Springer-Verlag, Berlin, 1996

\smallskip

\noindent[11] M. Horodecki, P. Horodecki and R. Horodecki, "Mixed-state 
entanglement and quantum communication" in {\it Quantum Information: An 
Introduction to Basic Theoretical Concepts and Experiments}, C. Alber 
{\it et al.} Eds., Springer-Verlag, Berlin, 2001    

\smallskip

\noindent[12]  L. Henderson, V. Vedral, {\it Classical, quantum, and total correlations}, 
arXiv:quant-ph/0105028, 8 May 2001

\smallskip

\noindent[13] M. Keyl, Physics Reports {\bf 369} (2002) 431 

\smallskip

\noindent[14] W.A. Majewski, {\it On entanglement of states and quantum 
correlations}, arXiv:math-ph/0202030, 21 Feb. 2002

\end{document}